\documentclass{mem}
\usepackage{natbib}\usepackage{txfonts}\usepackage{balance}
\usepackage{graphicx}
\usepackage[a4paper,breaklinks,dvipdfm]{hyperref}
\idline{75}{282}
\begin{document}
\def\teff{$T\rm_{eff }$}
\def\kms{$\mathrm {km s}^{-1}$}

\title{
Disc outbursts in various types of  binary systems
}

   \subtitle{}

\author{
J.-P. Lasota\inst{1,2} 
          }

  \offprints{J.-P. Lasota}

\institute{Institut d'Astrophysique de Paris, UMR 7095 CNRS, UPMC Univ Paris 06, 98bis Bd Arago, 75014 Paris, France
\and
Astronomical Observatory, Jagiellonian University, ul. Orla 171, 30-244 Krak\'ow, Poland
\email{lasota@iap.fr}
}

\authorrunning{Lasota}

\titlerunning{Disc outbursts}

\abstract{I discuss some aspects of the Disc Instability Model applied to outbursts in various types of  binary systems. I lament the general
lack of interest in the subject.

\keywords{accretion, accretion discs --  instabilities -- Stars: binaries: close -- 
Stars: dwarf novae -- X-rays: binaries -- X-rays: individual(ESO 243-49 HLX-1) -- Galaxies: star clusters }
}
\maketitle{}

\section{Introduction}

Disc outbursts are observed in all kind of binary systems. The best observed are outbursts of dwarf-nova stars where
the disc fed by a red-dwarf companion surrounds a white dwarf primary star. Initially, there was some confusion about  the 
dwarf-nova eruption site but \citet{smak71} clearly identified it as being the accretion disc and three years later \citet{osaki74}
suggested that dwarf nova outbursts are triggered by a disc instability. Few years later \citet{hoshi79} was the first to find
the origin of this instability\footnote{His paper singled out the convection as the main agent of the instability, while today
one would rather attribute it to the change in opacities, the two being obviously related} but it was the talk by Jim Pringle at the Sixth North American Workshop on Cataclysmic Variables, held at Santa Cruz in 1981 \citep[unpublished, but see][]{bapr82} that stimulated several authors \citep[][]{mm81,cannizzoetal82,flp83,mo83,smak84a}
to publish articles with elaborate models of unstable
dwarf nova discs. Soon after \citet{smak84b} published a review article on dwarf-nova outbursts which still contains almost
everything that one needs to know about the subject. The more recent review article by \citet{lasota01} can be considered
as just an extended commentary to Smak's 1984 paper; the only major difference being that it addresses also
the model of outbursting discs in X-ray binaries containing black holes or neutron stars. 

At present this 30 year old model, the Disc Instability Model (hereafter DIM) is generally believed to provide a correct description
of dwarf novae and X-ray transients. The word ``believe" is used here in its basic sense, i.e. ``accept that (something) is true, especially
without proof" (OED). Indeed, many counter-example to this believe are ignored by most of the interested community. Or one should rather
say: by the community that should be interested in these matters.

The DIM in its standard, or rather original, version assumes that, during the outburst cycle, the mass transfer from the secondary star, i.e. the
rate at which the disc is fed with matter, is constant. This simplifying assumption had also the advantage that it was clearly demarking the DIM
from its original rival: the Mass Transfer Instability (MTI) model that has been finally discarded, the mechanism able to trigger
it in the stellar companion remaining elusive. However, the assumption about the constancy of mass-transfer rate encounters two basic difficulties.
First, the model with this assumption produces regular, periodic outburst patterns that have never been observed. Second, real CVs
show fluctuations, sometimes very large, of the rate at which matter is transferred from the companion star. It is clear therefore that the
constant $\dot M$ assumption must be abandoned but this adds a free function of time to the model.

In addition the DIM is not able to reproduce the so-called superoutbursts, observed in SU UMa stars, whose amplitude is larger and duration longer than 
observed in normal outbursts.
Some systems, such as the celebrated short-period binary WZ Sge show superoutbursts only. To account for superoutbursts the DIM must be
substantially modified. Two main possibilities have been explored. The first consists in adding an additional source of viscosity: tidal torques
acting on the supposedly deformed eccentric disc. An important evidence in favour of this eccentricity is supposed to be the presence of so-called superhumps
modulated with a period slightly different from the orbital. The resulting increased viscosity causes a prolonged accretion episode resulting in large
amplitudes and long duration. This Tidal Thermal Instability model \citep[TTI; see][and references therein]{osaki05} has encountered several
serious difficulties when compared with observations \citep[see e.g.][]{B-MH02,hl05,hl06} but probably the strongest challenge to its application
to superoutbursts has been the paper by \citet{smak09} in which he showed that the presumed evidence of disc eccentricity during superoutbursts
resulted ``either from errors, or from arbitrary, incorrect assumptions". 
\begin{figure}[]
\resizebox{\hsize}{!}{\includegraphics[clip=true]{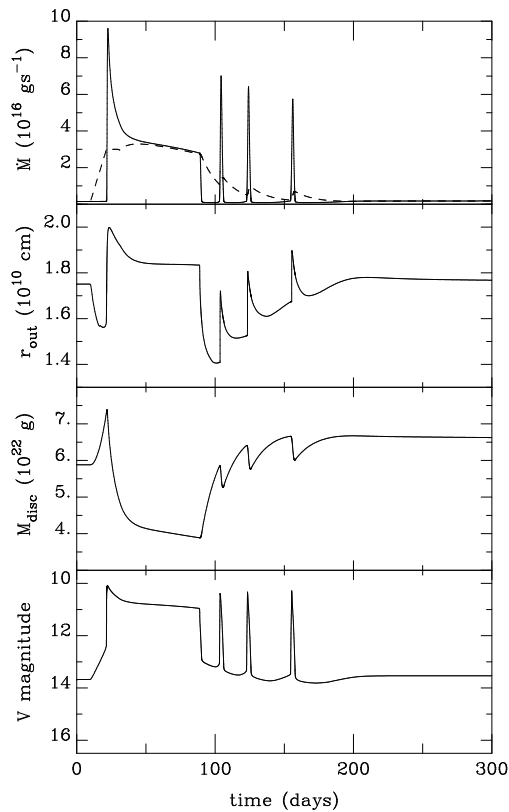}}
\caption{ \footnotesize Response of a steady accretion disc to a mass transfer rate variation by a factor 20. The parameters used are those of EG Cnc,
Irradiation is included as is the presence of a hole in the disc as well as the existence of a tidal instability. Initially $\dot M_2 = 1.5\times 10^{15} \rm g s^{?1}$; at
t = 10; $\dot M_2$ increases to $3.0 \times 10^{15} \rm g s^{?1}$, and at t = 30 returns to its quiescent value. The top panel shows the mass accretion
rate onto the white dwarf (solid line) and the mass transfer rate from the secondary (dashed line); the other panels show the
outer disc radius, its mass and visual magnitude. \citep[][]{B-MH02}
}
\label{fig:EGCnc}
\end{figure}
The second possibility of including superoutbursts into the DIM framework consists in explaining their amplitudes by an enhancement of the 
mass-transfer rate during the initial phases of an outburst that otherwise would have belonged to the normal category. This model has been
also proposed by \citet{osaki85} who, however, has quickly abandoned it in favour of the TTI model. \citet{hlh97} applied the enhanced
mass-transfer (EMT) model to WZ Sge whose fluence and very long (30 years) recurrence time (after the detection of its first outburst it
was considered to be a nova) could be reconciled with the DIM (also in its TTI version) only if the viscosity parameter $\alpha$ in quiescence was
assumed to be one or two orders of magnitude lower than in other dwarf novae. Not only the very long recurrence could be accounted for only
by lowering the value of $\alpha$ but also the mass accreted during the outburst could be explained only with very low values of the
viscosity parameter, the mass-transfer rate being too low to allow accumulating the required amount of mass despite the exceptionally long
recurrence (accumulation) time \citep{smak93}. \citet{B-MH02} combined the TTI and EMI models to reproduce the lightcurve of EG Cnc
(Fig. \ref{fig:EGCnc}).

\citet{smak08} proposed what he calls ``a purely observational scenario": \textsl{Superoutbursts are due and begin with a major enhancement in the mass transfer rate. During the "flat-top" part of the superoutburst the mass transfer rate decreases slowly, causing the observed luminosity to decline. The superoutburst ends when the mass transfer rate decreases below its critical value, resulting in a transition to the quiescent state of the dwarf nova cycle. }
The cause for this major enhancement is supposed to be irradiation of the secondary star by the accretion generated flux. If the regions close to the critical
$L_1$ point are screened by the disc this mechanism is unlikely to work \citep[][]{vh08}.  On the other hand,  \citet{smak11} found in five dwarf novae undergoing 
superoutbursts a direct evidence that irradiation of the companion is modulated at the superhump period thus supporting his interpretation of these features 
as being due to irradiation controlled mass transfer rate resulting in modulated dissipation of the kinetic energy of the stream.  It seems therefore
that irradiation controls mass-transfer during superoutbursts. One way solving the contradiction with \citet{vh08} is to consider the possibility of
a non-planar, warped disc.

\section{A short course in the DIM}
\label{sec2}
In a stationary (constant accretion rate), Keplerian accretion disc the effective temperature falls as $R^{3/4}$. This follows from
angular momentum and mass conservation only \citep[see e.g.][]{FKR02}.  The effective temperature at the inner disc edge can be written as
\begin{equation}
T_{{\rm in}}\approx6\times10^{7}\,\left(\frac{L_{38}}{\eta_{0.1}\, M_{1}^{2}}\right)^{1/4}\, x^{-3/4}\,{\rm K\,,
\label{tin}}
\end{equation}
 where $L_{38}$ is the luminosity in units of $10^{38}$ erg\,s$^{-1}$,
$\eta=0.1\eta_{0.1}$ is the accretion efficiency, $M_{1}$
accreting body mass in solar units and $x=c^{2}R/2GM$
is the radius measured in units of the Schwarzschild radius. $T_{{\rm in}}\approx 4\times 10^4$K for a cataclysmic variable
($L_{38}\sim 10^{-5}$, $\eta_{0.1}\sim 1.5\times 10^{-3}$, $x\sim 3.3 \times 10^3$) and $\sim 10^7$K for a neutron star 
or a stellar-mass black hole accreting close to the Eddington limit. In general discs around compact objects are hot
and fully ionized in their inner parts but if large enough they will reach temperatures at which hydrogen (or another dominant element)
starts recombining. The recombination process causes drastic changes in opacities modifying the disc cooling mechanism.
As a result the disc becomes thermally unstable. This is the essence of the DIM: a large enough hot stationary disc is always unstable.
For a given mass the critical size for instability depends on the accretion rate only:
\begin{equation}
R_{\rm crit}= 9 \times 10^4\,\left(\frac{\dot M/\dot M_{\rm Edd}}{M/M_{\odot}}\right)^{1/3}R_{S},
\label{eq:rcrits}
\end{equation}
where $\dot M_{\rm Edd}=  L_{\rm Edd}/0.1c^2$ is the Eddington accretion rate and $R_{S}=2GM/c^2$ the Schwarzschild radius.
For a given fraction of the Eddington accretion rate the instability radius descend deeper into the accreting body potential well. This
is the reason why, even when unstable, accretion discs around $> 10^8 M_{\odot}$ black holes do not produce dwarf-nova type
outbursts \citep[][]{hmv09}
\begin{figure}[]
\resizebox{\hsize}{!}{\includegraphics[clip=true]{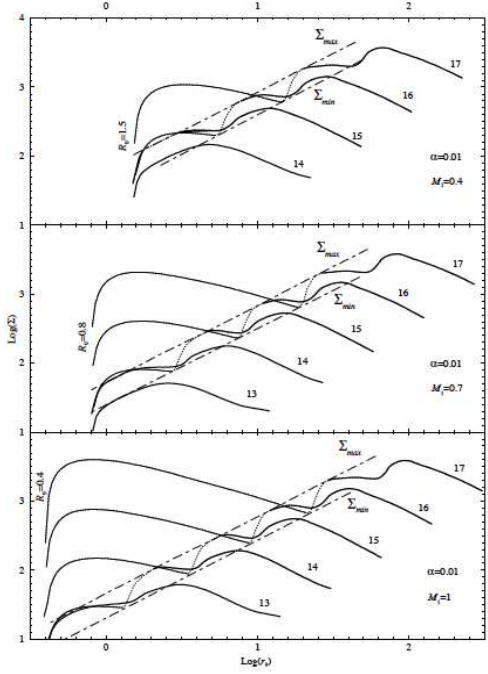}}
\caption{
\footnotesize
Surface density profiles ($\Sigma(r)$) for equilibrium discs around white dwarfs for three mass values:
0.4, 0.6 and 1.0 $M_{\odot}$
}
\label{fig:equilib}
\end{figure}
We will illustrate the principle of the DIM by considering a sequence of models of discs around white dwarfs. 
In Fig. \ref{fig:equilib} we represent stationary disc surface-density $\Sigma (R)$ profiles for three values of the central white-dwarf
mass $0.4,\, 0.6$ and $1.0M_{\odot}$ and five values of the accretion rate in the units of $\rm g/s$: $\log \dot M= 13, 14, 15, 16$ and 17.
The viscosity parameter is $\alpha=0.01$.
Each line labeled by a value of the accretion rate represents a stationary disc solution. The inner ``cut-off" is caused by the 
no-stress inner boundary condition $\nu \Sigma = 0$ and the number attached is the values of the white-dwarf radius since the disc
is assumed to extend down to the star's surface. Further away from the edge the column density falls like $R^{-3/4}$ 
\citep[see e.g.][]{FKR02} until it reaches a critical value marked as $\Sigma_{\rm min}$ where the slope changes and becomes
positive. In turn, this positive-slope branch encounters a second critical surface-density $\Sigma_{\rm max}$ where the slope
changes to negative. The more familiar, local picture of disc equilibria forming the celebrated S-curve can be recovered by 
considering all possible stationary solutions at a given radius. Taking a radius well inside the discs it is easy to see that the
$\dot M (\Sigma)$ points form an S-curve. In Fig. \ref{fig:equilib} going up from low $\Sigma$ and $\dot M$ one encounters
first points on the thick-line segments of the equilibrium curves -- they represent a growing segment of the $\dot M (\Sigma)$
curve (the ``lower branch" of the S-curve), then one gets to the dotted segments whose points will form a decreasing segment of
the $\dot M (\Sigma)$ curve (the ``middle branch") and finally points collected on the thin continuous lines will form what
is known as the upper branch of the S-curve.
\begin{figure}[]
\resizebox{\hsize}{!}{\includegraphics[clip=true]{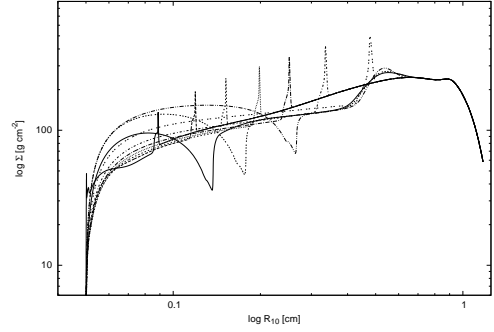}}
\caption{ \footnotesize Heating and cooling fronts in a helium dominated $Y=0.98, Z=0.02$ accretion disc. (Figure courtesy of Iwona Kotko).
}
\label{fig:profile}
\end{figure}
Therefore negative slopes in $\Sigma(R)$ correspond to stable solutions while a solution with surface density increasing with radius is
unstable. At given radius this corresponds to the familiar stability/instability $\partial \dot M/\partial \Sigma > 0/ 
\partial \dot M/\partial \Sigma < 0$  conditions.

Hence for a given mass-transfer rate the stability properties of an accretion disc depend on it size: if its outer radius $R_D > R_{\rm crit}$
no stable equilibrium configuration exists\footnote{Which has not prevented some astronomers from attempting to fit stationary
disc spectra assuming effective temperatures spanning from 30 000 to 4000 K.}. In such cases the disc goes into an outburst cycle in a hopeless quest for a stable solution. For
a white dwarf primary this results in a dwarf-nova outburst cycle.

The disc is brought up to a hot bright state by a heating front propagating out- or inwards depending on the mass-transfer rate (in Fig. \ref{fig:profile} the
heating front propagates outwards). This phase represents the rise to maximum. Propagating through the disc, the heating front redistributes\footnote{The front propagation is a thermal {\sl and} viscous process.} the temperature
(flat profile at quiescence) and surface-density (at quiescence increasing with radius) distributions so that at peak luminosity the disc has quasi-stationary structure
with $\dot M(R) \sim \rm const.$ and $T_{\rm eff}\sim \Sigma \sim R^{-3/4}$. Since at maximum $\dot M \approx \dot M_{\rm crit}^+(R_D)$ -- the upper
critical accretion rate -- is by construction (the disc is unstable) higher than the mass-transfer rate, the disc must empty out. This occurs on a viscous time-scale:
$\sim R^2/\nu \sim R^2/\alpha c_s H$ \citep[using the][prescription for the kinematic viscosity
coefficient $\nu=\alpha c_{s}H$, with $H$ the discs's semi-scale-height]{ss73}. Thus by observing dwarf-nova decay from maximum one can (measuring the disc's size and estimating its temperature -- both procedures rather straightforward, especially in eclipsing systems) determine the viscosity parameter $\alpha$. Analyzing dwarf nova light-curves
collected by \citet{vp83}, \citet{smak99} has determined that $\alpha \approx 0.2$ in hot, ionized accretion discs. It is worth stressing that MHD simulations obtain, at best, 
values an order of magnitude lower \citep[see e.g.][]{sorathiaetal11}. The ball is clearly in the simulators court.

The viscous decay from maximum proceeds through  a sequence of quasi-stationary configurations. This pushes the outer disc regions into the unstable regime.
As a result a cooling front forms bringing the outer disc into the cold quiescent state. During the cooling front propagation one recovers (for obvious reasons) a global 
disc structure similar to that when discussing above the stability of equilibrium discs  (compare Fig. \ref{fig:profile} with Fig. \ref{fig:equilib}). The cooling
front brings the disc to a cold state (onto the ``cold branch" of the S-curve) but (contrary to what is sometimes asserted) does not govern the decay from maximum.
The reason being that the speed of the cooling front is (slightly) smaller than the viscous speed ($\sim \nu/R$) in the hot discs it penetrates into \citep[][]{mns99}.

It can be considered that the DIM is rather successful in describing the rise to and the decay from maximum. The same cannot be said of the quiescent phase
of the dwarf nova outburst cycle \citep[see e.g.][for a discussion]{lasota01}.

\subsection{SS Cygni is not a dwarf nova (but looks like one)}

The cataclysmic variable star SS Cygni is the brightest and best observed dwarf-nova. Although formally not the prototype of its subclass (it is of U Geminorum type) it is in fact the epitome of the dwarf nova undergoing normal outbursts.  The reproduction by the DIM of the two main types of normal outbursts observed in SS Cyg (``A" and ``B" or ``long" and ``short") might require small modulations of the mass-transfer rate but the multi-wavelength long-term lightcurve of SS Cyg is quite well reproduce by the standard DIM \citep[][]{shl03} if the disc is truncated \textsl{and} the distance is not larger than $\sim 100$\,pc. However, the HST/FGS parallax distance to SS Cyg has been determined to be $166\pm 12$\,pc, a result confirmed by the modeling of this system's properties by \citet{bitneretal07} whose results are consistent with a distance of $\sim 140-170$\,pc. \cite{sl07} showed that at such distance both the accretion rate during outburst \textsl{and} the mean mass-transfer rate is too high to be compatible with the DIM. With such a mass-transfer rate, independent of the DIM, SS Cyg should be a nova-like star and show no outbursts. In other words: at 166 pc SS Cyg is not a dwarf nova.

Recently, \cite{smak10} has challenged part of this conclusion by deriving the mass-transfer rate from the hot-spot (where the accretion stream from the companion encounters the outer disc rim) luminosity. His result is 150 lower than that of \citet{sl07} and satisfies the DIM criterion. However, since in this new framework the accretion rate in outburst is still the one determined by \citet{sl07}, explaining the outburst by the DIM require enhancing the mass-transfer rate {\sl during outburst} by the same factor of 150.  \cite{smak10} concludes that: ``Nothing is wrong with SS Cyg, nor with theory of dwarf-nova outbursts". Although it is difficult to disagree with the first part of this statement, the truth of second is less obvious. Indeed, enhancing the mass-transfer rate by a factor 150 is a highly non-trivial task. Explaining superoutbursts of SU UMa and WZ Sge stars requires a factor 100 ``only" \citep[][]{hlh97,smak04} but outbursts of SS Cyg do not look at all like superoutbursts. As mentioned above, the mechanism of the putative enhancements is rather uncertain but e.g.  Smak (2004b) estimated that irradiation will not work for orbital periods longer that $\sim 6$\,hr which is rather unfortunate since it is just the orbital period of SS Cyg (6.6\,hr). 

\cite{smak10} explains the visual magnitude by a mass-transfer rate $\dot M\approx 6.3 \times 10^{16}$g\,s$^{-1}$. But this is only a small fraction (0.047$\pm$0.005) of the total quiescent visual flux. According to \citet{bitneretal07} a fraction $f_{d}=0.535\pm0.075$ of this is emitted by the accretion discs. In the framework of the DIM the properties of this (non-stationary) disc are rather well constrained. Its effective temperature cannot be higher than the lower critical temperature $T_{\rm crit}^-=5210\, R_{10}^{-0.1}M_1^{0.04}$ \citep[][]{ldk08} and the (increasing with radius) accretion rate should be everywhere lower than
\begin{equation}
\dot M_{\rm crit}^-=2.64\times 10^{15}\, \alpha^{0.01}_{0.1}R_{10}^{2.58}M_1^{-0.85}. 
\end{equation}
The disc radius $R_{10}$ is in units of $10^{10}$\,cm. 

Let us assume for simplicity that half of the disc visual flux of $V=5.6$ is emitted by the disc in quiescence. This corresponds to a luminosity in $V$:
\begin{equation}
L_V\approx 1.0 \times 10^{33}\,\rm erg\,s^{-1}
\label{lv}
\end{equation}
One can estimate the disc effective temperature corresponding to this luminosity by assuming that it is roughly constant with radius as shown by observations of eclipsing dwarf-novae and required by the DIM. Using the system parameters determined by \citet{bitneretal07} one gets for the outer disc radius ($R_D=0.9\,R_1$) $4.7 - 5.5\times 10^{10}$\,cm \citep[see][]{sl07}. $R_1$ is the distance to the $L_1$ point.
Therefore the effective temperature is $T_{\rm eff}\approx 6540$ K. This is of course higher than the critical temperature. With such a temperature the disc in SS Cyg cannot be in a quiet cold state as required by the DIM. Let us note that the value we derived is certainly a {\sl lower limit}, since radii of quiescent discs in dwarf-novae are smaller than $0.9R_1$ \citep[see e.g.][]{H-AW96} and we took into account only the V luminosity.

Using more refined methods Smak (private communication) showed that one obtains $T_{\rm eff}$'s lower than critical by taking $f_{d + 1} \leq 0.45$ (where $f_{d + 1}$
is the luminosity fraction emitted by the disc plus the primary). In view of the uncertain value of this parameter \citep[][]{bitneretal07} this is not an unacceptable modification but (because
of flux conservation) it implies a higher luminosity of the secondary making it K2 or earlier in contradiction with the K4.5 in \citet{bitneretal07}. In other words, at 166 pc there is too much flux in SS Cyg to accommodate both the DIM and the observed properties of this binary's components. Clearly the distance-to-SS Cyg problem requires serious investigations, after all it is the prototype dwarf nova, but nobody (except for Smak and the present author) seem to be interested. So much for the \textsl{Golden Age of Cataclysmic Variables}.

\section{Outbursts of the hyperluminous X-ray source HLX-1 in ESO 243-49}

\begin{figure}[]
\resizebox{\hsize}{!}{\includegraphics[clip=true]{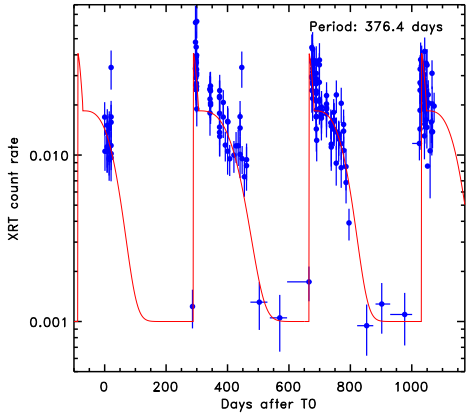}}
\caption{ \footnotesize The lightcurve of HLX-1. The red line represents a fit with two exponentials with decay times $\tau_1= 20 \pm 4$ days,
$\tau_2= 130 \pm 3$ days; the second outburst is longer than the three other ones with $\tau_1= 110 \pm 4$ days, $\tau_2= 135 \pm 3$ days.
The recurrence time is $\sim 376$ days. (Figure courtesy of Didier Barret).
}
\label{fig:lchlx1}
\end{figure}

Observing a Fast Rising Exponentially Decaying (FRED) lightcurve of a newly discovered X-source one is immediately tempted to attribute it
to the DIM. This was the case of HLX-1 in the halo of the edge-on S0a-type galaxy ESO 243-49 which is the brightest known ultraluminous X-ray source 
\citep[ULX; see][for a review]{roberts07} with a maximum luminosity $>$10$^{42}$ erg s$^{-1}$ \citep{Farrell_nature,godetetal09}. This
has not only the highest (by an order of magnitude) luminosity ULX but is also unique in showing clear, large amplitude outbursts \citep[see Fig. \ref{fig:lchlx1}; and][]{godetetal11} whose spectral
behaviour is very similar to that observed in stellar-mass Black Hole X-ray Binary transients.
The luminosity of HLX-1 ($\sim10^{42}$ erg s$^{-1}$) and the variability
timescale ($\sim10^{7}$ s) imply a huge accretion rate onto the compact
object, of the order of $10^{-4}$ M$_{\odot}$ yr$^{-1}$. Sustaining
such a high mass transfer rate excludes wind accretion and implies the presence of a stellar companion 
filling its Roche-lobe (if only during part of its orbit) losing matter
that forms an accretion disc around the black hole. With such an assumption one encounters a fundamental
difficulty because for a black hole mass assumed to be around $10^{4}M_{\odot}$
the mass ratio $q=M_*/M_{\rm bh}$, where $M_{\rm bh}$ and $M_*$ are respectively
the black-hole and stellar-companion masses, will typically be $q\approx10^{-4}-10^{-3}\ll1$.
For such a small $q$, matter circularizes very close to the donor
star and the 2:1 Lindblad resonance appears at a radius $\approx0.63a$ (where $a$ is the orbital separation) 
within the primary's Roche lobe \citep{lin}, affecting the accretion
disc formation and structure.
The only thing that is certain is that the standard formulae for the sizes of the elements of the binary cannot
be used in such a case. For lack of better solution one assumes the disc forms and has a size $R_{{\rm D}}\approx a$.

As explained in Sect. \ref{sec2}, to be unstable the disc must have a radius larger than the critical value of Eq. (\ref{eq:rcrits}), hence for HLX-1 with 10$^{4}$ M$_{\odot}$ and
$\dot M=10\,\dot M_{\rm Edd}$ ($L_{\rm max}\approx L_{\rm Edd}$) the condition is: 
\begin{equation}
R_D>R_{\rm crit}= 2.7 \times 10^{13}M_4^{-1/3} \dot m_{10}^{1/3}\, \rm cm
\label{rcrithlx}
\end{equation}
where $M_4$ is the mass in units of M$_{\odot}$ and $\dot m_{10}=(10\,\dot M/\dot M_{\rm Edd})$. (The value given by Eq. (\ref{rcrithlx}) is slightly
different from that of Eq. (5) in \citet{lasotaetal11} because of different forms of criteria used -- the formulae for critical quantities are obtained from fits
to equilibrium S-curves and often not very exact.) The actual disc can be much larger than this limit because: i.) the heating front bringing the disc to maximum
luminosity could have stopped before reaching the outer edge; ii) in Eq. (\ref{rcrithlx}) we have assumed a non-irradiated disc whereas the disc in HLX-1 is
certainly irradiated and therefore stabilized \citep[see e.g.][]{dubusetal99}. Hence a larger critical radius.

At first sight there seem to be nothing wrong about the radius required by Eq. (\ref{rcrithlx}). As shown in \citet{lasotaetal11} the Roche geometry implies a
mean density of the companion star $\bar{\rho}\approx 0.057\, P_{d}^{-2}\,{\rm g\, cm^{-3}}$. From Eq. (\ref{rcrithlx}) it follows that the orbital
period is  $\ga$ 23 days. The secondary star mean density of this hypothetical companion is  $10^{-4}\ {\rm g\, cm^{-3}}$ implying
a red giant or a massive supergiant -- a very reasonable option.

However, even a quick look at the timescales completely shatters this optimistic conclusion. Indeed, the typical outburst timescale will
be linked to the viscous timescale of the minimum critical radius
\begin{equation}
t_{{\rm vis}}=\frac{R^{2}}{\nu}\approx115\ \alpha^{-1}T_{4}^{-1}R_{13}^{1/2}M_{4}^{1/2}{\rm years},
\label{tvis}
\end{equation}
 $T_{4}$ is the disc temperature in units of 10$^{4}$ K (we assume for simplicity that it is the temperature at $R_{\rm crit}$).
 Therefore the variability timescale is much too long for any reasonable
set of parameters. The variability in HLX-1 cannot be
related to the processes described by the DIM since that it is impossible to reconcile
these timescales with a disc large enough that its temperature at the
outer radius is lower than $\approx10^{4}$ K. Therefore the accretion disc in HLX-1 must be hot and thermally stable.

It seems that the only serious option left is that rejected for dwarf novae: the MTI. But that's another story \citep[][]{lasotaetal11}.

\section{Conclusions}

The Disc Instability Model in its pure, original form describes no real system. Even the prototype dwarf nova star, SS Cyg can be described only if its inner disc is truncated and 
the mass-transfer vary. If its Hubble parallax gives the true distance then it cannot be described by the DIM at all. It should rather be nova-like star, but obviously it is not. The lack
of interest in this fundamental problem is rather surprising.
Although the DIM has been applied, more or less successfully, to outbursts of X-ray transient binaries, rather surprisingly it cannot be applied to outbursts of the
Intermediate Mass Black Hole HLX-1 despite them exhibiting most of the required properties.

\begin{acknowledgements}
Didier Barret, Olivier Godet and Iwona Kotko are thanked for their help in preparing this article. Joe Smak has inspired a lot of my research in the subject, especially by his critical but always friendly remarks. I am grateful to Franco Giovannelli for kindly inviting me to the Montello conference giving me the opportunity to publish the present complaint. This work has been supported in part by the French Space Agency CNES.
\end{acknowledgements}

\bibliographystyle{aa}

\end{document}